\documentclass[11pt]{article}
\usepackage[pdftex]{graphicx}
\usepackage{graphics}
\usepackage[top=1cm,bottom=2cm,right=2cm,left=2cm]{geometry}
\begin{document}
\begin{center}{\Large \bf Dynamics of the localized  nonlinear waves in  spin-1  Bose-Einstein condensates with time-space modulation}
\end{center}
%===============================作者姓名,地址===================
\begin{center}
{\it Yu-Qin Yao$^{1}\footnote{e-mail: yyqinw@126.com }$, Wei Han$^{2}$, Ji Li$^{3}$ , Hui-lan Wu$^{1}$ and Wu-Ming Liu$^{3} \footnote{e-mail: wliu@iphy.ac.cn }$ }
\end{center}
\begin{center}{\small \it $^{1}$ Department of
 Applied Mathematics, China Agricultural University, Beijing 100083, People's Republic of China\\
 $^{2}$ Key Laboratory of Time and Frequency Primary Standards,
National Time Service Center, Chinese Academy of Sciences, Xi’an 710600, People's Republic of China \\
 $^{3}$ Beijing National Laboratory for Condensed Matter Physics, Institute of Physics, Chinese Academy of Sciences, Beijing 100190, People's Republic of  China}
\end{center}
\vskip 12pt { \small\noindent\bf Abstract }{ }
We investigate the dynamics of the  localized nonlinear matter wave in spin-1  Bose-Einstein condensates  with trapping potentials and nonlinearities dependent on time and space. We solve the three coupled Gross-Pitaevskii equation by similarity transformation and obtain two families of exact matter wave solutions in terms of Jacobi elliptic functions and Mathieu equation. The localized states of the spinor matter wave  describe the dynamics of vector
breathing solitons, moving breathing solitons, quasibreathing solitons and resonant solitons. The results of  stability  show that  one order vector breathing solitons, quasibreathing solitons, resonant solitons,  and the moving breathing solitons $\psi_{\pm1}$ are all stable but the moving breathing solitons $\psi_0$ is unstable. We also present the experimental parameters to realize these phenomena in the future experiments.
%{\small\noindent\bf PACS number(s):} {05.45.Yv, 67.85.-d, 03.75.Lm, 42.65.Tg} \vskip 6pt

\centerline{\bf I. INTRODUCTION}
%\section{Introduction}
For a decade, the experimental realization of Bose-Einstein condensations (BECs)  at ultra-low temperatures has attracted great interest in the atomic physics communication \cite{pc,ro}. In recent years, one of the most important developments in BECs was the study of the spinor condensations. The idea of the spinor condensations was proposed by Ho and Ohmi \cite{ho,oh}. Stamper-Kurn et al. created the spinor condensations in experiment \cite{sk}, which provides a new perspective to observe phenomena that are not present in single-component BECs. These included the formation of spin domains and spin textures \cite{js,mv}.
Later, the spinor BECs with $F>1$ have also been studied theoretically \cite{hs,hsm}. In contrast to single and two-component BECs, the spin-F condensates described by macroscopic wave functions with 2F+1 components have some new characteristics, including the vector character of the order parameter and the changed role of the spin relaxation collisions.
Here we focus on BECs of alkali atoms in the F=1 hyperfine state, such as $^{23}Na$,
$^{87}Rb$  and $^{7}Li$ \cite{js,hj}, restricted to one-dimensional space by purely optical means. In the absence of an external magnetic fields, the three internal states $m_F=1,0,-1$, with $m_F$ the magnetic quantum number, are generated, in which an $m_F=1$ and an $m_F=-1$ atom can collide and produce two
 $m_F=0$ atoms and vice versa.  Under the mean-field approximation, the dynamics of the spinor condensates can be described by three-component Gross-Pitaevskii equation (GPE). Stimulated by the experiments done by JILA \cite{jila} and MIT \cite{mit} groups and the theoretical works of Ho, Ohmi and Machida \cite{hoo,mach}, many studies have been done [15-23].
%\cite{cv,mk,ds,ac,ji,mu,ll,bj,puh}.
While, there is few work on the spinor BECs with time-space modulation to present.

   Matter waves as the natural outcomes of the mean-field descriptions
have been observed experimentally and investigated theoretically \cite{zd,ke,be}. For example, matter wave solitons in atom optics could be used for
  applications in atom laser, atom interferometry and coherent atom transport. Moreover, it is also helpful to the realization of quantum information processing and computation \cite{optic}. So it is interesting to develop a new technique for constructing particular solitons. One possible technique is to alter the interatomic interactions by means of external magnetic fields.  Recent experiments show that the effective scattering length can be tuned by Feschbach resonance \cite{si,bam,mt}. This  brought about a good  proposal for manipulation of the nonlinear excitations and matter wave by controlling the time-dependent or space-dependent  scattering strength [31-34].
% \cite{jb,ry,vvk,ds2}.
In Refs. \cite{liangzx,ekeng}, the matter wave solitons in BECs with time-dependent scattering length was investigated. Nonlinear matter wave in atomic-molecular BECs with space-modulated nonlinearity and in two-component BECs with time-space modulation nonlinearities have been studied.\cite{yao,at}.

In this paper, we consider the spin-1 BECs with space and time dependent nonlinearities and trapping potentials, which can be described by a system of three-component GPEs. Different from the one-, two-component BECs, we can use an optically induced Feshbach resonance \cite{jmg} or a confinement induced resonance \cite{tbe} to tune the nonlinearities in the spinor BECs.
 Two kinds of localized nonlinear matter wave  are given based on the Mathieu equation and Jacobi elliptic function, which take the form of vector solitons. We investigate in detail
the vector breathing solitons, moving breathing solitons, quasibreathing solitons and resonant solitons. Dynamical stability of the obtained vector solitons are studied by means of the numerical simulations  and the global stability of the different types of vector solitons  are analysed.  The results show that all the one order vector breathing solitons, quasibreathing solitons and resonant solitons  are stable,  as for the moving breathing solitons,   the matter wave $\psi_1$ and $\psi_{-1}$ are stable but   $\psi_0$ is unstable.

The paper is organized as follows. In Sec.II, the localized nonlinear wave solutions are presented  based on the Mathieu equation and Jacobi elliptic function. Four kinds of vector solitons, including vector breathing solitons, moving breathing solitons, quasibreathing solitons and resonant solitons, are illustrated in Sec.III. Sec.IV  analyse the dynamic stability of the vector soliton by the numerical simulations. An conclusion is given in the last section.

\vskip 15pt
%\section{Localized nonlinear matter wave solutions}
\centerline{\bf II. LOCALIZED NONLINEAR MATTER WAVE SOLUTIONS}
We consider the spin-1 BECs confined in the trapping potential $V_{ext}=\frac{m}{2}(\omega_x^{2}x^2+\omega_{\perp}^{2}(y^2+z^2))$ with $m$  the mass of $^{23}Na$ atoms, $\omega_x$ and $\omega_{\perp}$ are the confining frequencies in the transverse and axial directions.
The spin-1 BECs can be described by vectorial wave function $\Psi(x,t)=(\psi_1(x,t),\psi_0(x,t),\psi_{-1}(x,t))^{T}$ with the components corresponding to the three values of the vertical spin projection $m_F=+1,0,-1$. When the temperature is lower than the  critical temperature, the wave functions are governed by a set of three coupled dimensionless Gross-Pitaevskii equation \cite{ho,ac,ckl}
{\small\begin{equation}
\label{eqns:eq}
       \begin{array}{ll}
       i\frac{\partial \psi_1}{\partial t}=(-\frac{\nabla^2 }{2}+g_n(|\psi_1|^2+|\psi_0|^2+|\psi_{-1}|^2)
       +g_{s}(|\psi_1|^2+|\psi_0|^2-|\psi_{-1}|^2)+V(x)+E_1)\psi_1+g_s\psi_{-1}^{*}\psi_{0}^{2} , \\
        i\frac{\partial \psi_0}{\partial t}=(-\frac{\nabla^2 }{2}+g_n(|\psi_1|^2+|\psi_0|^2+|\psi_{-1}|^2)
       +g_{s}(|\psi_1|^2+|\psi_{-1}|^2)+V(x)+E_0)\psi_0+2g_s\psi_{-1}\psi_{0}^{*}\psi_1 , \\
      i\frac{\partial \psi_{-1}}{\partial t}=(-\frac{\nabla^2 }{2}+g_n(|\psi_1|^2+|\psi_0|^2+|\psi_{-1}|^2)
       +g_{s}(|\psi_{-1}|^2+|\psi_0|^2-|\psi_{1}|^2)+V(x)+E_{-1})\psi_{-1}+g_s\psi_{1}^{*}\psi_{0}^{2} . \\
       \end{array}
   \end{equation}}
Here, $\nabla^2=\frac{\partial^2}{\partial
x^2},$ $V(x)=\frac{ \omega^2 x^2}{2}$  is the external trapping potential,   $E_{j}\in R$ is the dimensionless Zeeman energy of spin component $m_F=-1,0,1$ and $|\psi_1|^2+|\psi_0|^2+|\psi_{-1}|^2$ is the total condensate density. The strength of the interaction is given by $g_n=\frac{4\pi \hbar^2 (a_0+2a_2)}{3 m},~g_s=\frac{4\pi \hbar^2(a_2-a_0)}{3 m}$ , where $\hbar$ is the reduced Planck constant  and  $a_0,~a_2$ are the s-wave scattering lengths scattering channel of total hyperfine spin-0 and spin-2, respectively \cite{kwm}. The length and time are measured in the unit of $\sqrt{\frac{\hbar}{m w_{\perp}}}$ and $w_{\perp}^{-1}$, respectively. Here we provide the experimental parameters for producing the spinor condensate composed of $^{23}Na$ \cite{nnk,bjd} with the total number  $N=3\times10^{6}$. The external trapping potentials is given by  $V=\frac{\omega x^2}{2}$
with $\omega=2\pi \times 230 Hz$. The scattering lengths $a_0=50 a_B$ and $a_2=55 a_B$ with Bohr radius $a_B=0.529 {\AA}$. In Refs. \cite{ua1,ua2} , the spinor BECS was illustrated systematically from experimental and theoretical progress. In our case, the scattering lengths depend on time and space, that is to say $g_n=g_n(x,t),g_s=g_s(x,t)$, which can be realized by controlling the induced Feschbach resonances optically or confinement induced resonances in the real BEC  experiments.

   In the following, we seek the exact localized solutions of   (\ref{eqns:eq}) for $lim_{|x|\rightarrow\infty}\psi_{\pm1,0}=0$.
  To do this,  the similarity transformation
  $$\Psi_1=\beta_1(x,t)U(X(x,t))e^{i\alpha_1(x,t)}$$
\begin{equation}
\label{eqns:st} \Psi_{-1}=\beta_{-1}(x,t)V(X(x,t))e^{i\alpha_{-1}(x,t)},
\end{equation}
$$\Psi_0=\beta_0(x,t)W(X(x,t))e^{i\frac{\alpha_1(x,t)+\alpha_{-1}(x,t)}{2}},$$
are taken  to transform  (\ref{eqns:eq}) to the ordinary differential
equations (ODEs)
 \begin{equation}
\label{eqns:ode}
       \begin{array}{ll}
   U_{XX}+b_{11}U^3+b_{12}UV^2+b_{13}W^{2}U=0 , \\
    V_{XX}+b_{21}U^2V+b_{22}V^3+b_{23}W^2V=0,\\
   W_{XX}+b_{31}V^2W+b_{32}U^2W+b_{33}W^3+b_{34}UVW=0,
       \end{array}
   \end{equation}
where $b_{ij},i=1,2,~j=1,2,3,4$ are constants. Substituting
(\ref{eqns:st}) into (\ref{eqns:eq}) and letting $U(X),~V(X),~W(X)$ to
satisfy (\ref{eqns:ode}), we obtain a set of partial differential
equations (PDEs). Solving this set of PDEs, we have
\begin{equation}
\label{eqns:al}\begin{array}{ll}
\alpha_{\pm 1}(x,t)=-\frac{\lambda(t)_tx^{2}}{2\lambda(t)}+\xi_{\pm 1}(t)-\frac{\delta(t)_t x}{\lambda(t)},~
\beta_{j}(x,t)=\theta_j(t)\lambda(t)^{-\frac{1}{2}}e^{\frac{(\lambda(t) x+\delta(t))^{2}}{2}},~ X=\frac{1}{2}\sqrt{\pi}erf(\lambda(t) x+\delta(t)),
   \end{array} \end{equation}
   $$g_n=-\frac{b_{12}\beta_{1}(x,t)^{2}+b_{11}\beta_{2}(x,t)^{2}}{4\beta_{1}(x,t)^{2}\beta_{2}(x,t)^{2}},~g_s=\frac{b_{12}\beta_{1}(x,t)^{2}
   -b_{11}\beta_{2}(x,t)^{2}}{4\beta_{1}(x,t)^{2}\beta_{2}(x,t)^{2}}$$
where
$erf(s)=\frac{2}{\sqrt{\pi}}\int_0^se^{-\tau^2}d\tau$ is
called an error function, the undetermined functions $\xi_{\pm 1}(t)$ and $\theta_j(t),~(j=-1,0,1)$  satisfy
\begin{equation}
\label{eqns:eqs}\begin{array}{ll}
\lambda\lambda_{tt}-2\lambda_t^{2}-(3\omega_2^{2}+\omega_1^{2})\lambda^{2}+\lambda^{6}=0,\\
-4\lambda_t\delta_t+2\delta_{tt}\lambda+2\lambda^{5}\delta=0,\\
-\delta_t^{2}+\delta^{2}\lambda^{4}-2\lambda^{2}\xi_{1t}-2E_1\lambda^{2}+\lambda^{4}=0,\\
-\delta_t^{2}+\delta^{2}\lambda^{4}-2\lambda^{2}\xi_{-1t}-2E_{-1}\lambda^{2}+\lambda^{4}=0,\\
-\delta_t^{2}+\delta^{2}\lambda^{4}-\lambda^{2}\xi_{1t}-\lambda^{2}\xi_{-1t}-2E_0\lambda^{2}+\lambda^{4}=0,\\
\theta_{jt}\lambda-\theta_j\lambda_t=0,~j=-1,0,1.
   \end{array} \end{equation}
 To obtain explicit solutions, we choose $\omega^2$ as the form
\begin{equation}
\label{eqns:w}
\omega^2=\omega_0^2+\epsilon cos(\overline{\omega}t),
 \end{equation} where $\epsilon \in(-1,1)$ and $\omega_0,~\overline{\omega}\in R$.
Now under the condition $E_1+E_{-1}=E_0$, solving (\ref{eqns:eqs}) gives
\begin{equation}
\label{eqns:ps}\begin{array}{ll}
\lambda(t)=(A sin^{2}\omega t+B cos^{2}\omega t+\frac{2\sqrt{\omega^{2}(AB\omega^{2}-1) }sin\omega t cos\omega t
}{\omega^{2}})^{-\frac{1}{2}},\\
\theta(t)_j=c_j \lambda,~j=-1,0,1,\\
\xi(t)_{\pm1}=\frac{1}{2}\int\frac{\lambda^4+\lambda^4\delta^2-\delta_t^{2}}{\lambda^2}dt-E_{\pm1} t, \\
\delta(t)=c_2e^{i\int\lambda^{2}dt}+c_3e^{-i\int\lambda^{2}dt},\\
   \end{array} \end{equation}
  % $$\delta=cos(arctan(\frac{A\omega^{2}tan\omega t+\sqrt{\omega^{2}(AB\omega^{2}-1)}}{\omega}))$$
where $c_j,~(j=-1,0,1,2,3)$ are arbitrary constants, and $A,~B$ are constants satisfying $AB-C^{2}=\frac{1}{\xi_1\xi_{1t}-\xi_2\xi_{2t}}$ with $(\xi_1,\xi_2)$
being two linear independent solutions of the Mathieu equation
\begin{equation}
\label{eqns:me}\xi_{tt}+\omega^{2} \xi=0.\end{equation}

When the interactions are attractive,  the ODEs (\ref{eqns:ode}) have two families of exact solutions
\begin{equation}
\label{eqns:sl}\begin{array}{ll}
U^{(1)}=-\sqrt{\frac{-c_0^{2}b_{12}+2c_{-1}^{2}\mu_1^{2}}{2c_1^2b_{12}}}cn(\mu_1 X,\frac{\sqrt{2}}{2}),\\
V^{(1)}=\sqrt{\frac{-c_0^{2}b_{12}+2c_{-1}^{2}\mu_1^{2}}{2c_{-1}^{2}b_{12}}}cn(\mu_1 X,\frac{\sqrt{2}}{2}),\\
W^{(1)}=cn(\mu_1 X,\frac{\sqrt{2}}{2}),
   \end{array} \end{equation}

%\begin{equation}
%\label{eqns:s2}\begin{array}{ll}
%U^{(2)}=-\sqrt{\frac{-c_0^{2}b_{12}+4c_{-1}^{2}\mu_2^{2}}{2c_1^2b_{12}}}sn(\mu_2 X,\frac{\sqrt{2}}{2}),\\
%V^{(2)}=\sqrt{\frac{-c_0^{2}b_{12}+4c_{-1}^{2}\mu_2^{2}}{2c_{-1}^{2}b_{12}}}sn(\mu_2 X,\frac{\sqrt{2}}{2}),\\
%W^{(2)}=sn(\mu_2 X,\frac{\sqrt{2}}{2}),
 %  \end{array} \end{equation}
and
\begin{equation}
\label{eqns:s3}\begin{array}{ll}
U^{(2)}=-\sqrt{\frac{-c_0^{2}b_{12}+c_{-1}^{2}\mu_2^{2}}{2c_1^2b_{12}}}sd(\mu_2 X,\frac{\sqrt{2}}{2}),\\
V^{(2)}=\sqrt{\frac{c_0^{2}b_{12}+c_{-1}^{2}\mu_2^{2}}{2c_{-1}^{2}b_{12}}}sd(\mu_2 X,\frac{\sqrt{2}}{2}),\\
W^{(2)}=sd(\mu_2 X,\frac{\sqrt{2}}{2}).
   \end{array} \end{equation}
Here, $\mu_1,~\mu_2$ are arbitrary constants, $cn,~sd=sn/dn$ are Jacobi elliptic functions. When the interactions are repulsive, the solutions of the  the ODEs (\ref{eqns:ode}) are similar to (\ref{eqns:sl}) and (\ref{eqns:s3}).
When imposing the bounded condition $lim_{|x|\rightarrow\infty}\psi_{\pm1,0}(x)=0$, we have $\mu_1=\frac{2(2n+1)K(\frac{\sqrt{2}}{2})}{\sqrt{\pi}}$ and $\mu_2=\frac{4m K(\frac{\sqrt{2}}{2})}{\sqrt{\pi}}$,
where the natural number $n$  and $m$  are the order of the solitons, and $K(\frac{\sqrt{2}}{2})=\int_0^{\pi/2}[1-\frac{1}{2} sin^2\xi]^{-\frac{1}{2}}d\xi.$

Based on (\ref{eqns:al}),~(\ref{eqns:ps})-(\ref{eqns:s3}), we work out two families of exact solutions of the dimensionless Gross-Pitaevskii equation  (\ref{eqns:eq})
\begin{equation}
\label{eqns:fs}\begin{array}{ll}
\psi_1^{(j)}=c_1\sqrt{\lambda(t)}e^\frac{\delta(t)^2}{2}e^{\gamma(x,t)+i\alpha_1(x,t)}U^{(j)}(X),\\
\psi_{-1}^{(j)}=c_{-1}\sqrt{\lambda(t)}e^\frac{\delta(t)^2}{2}e^{\gamma(x,t)+i\alpha_{-1}(x,t)}V^{(j)}(X),\\
\psi_0^{(j)}=c_0\sqrt{\lambda(t)}e^\frac{\delta(t)^2}{2}e^{\gamma(x,t)+\frac{(i\alpha_1(x,t)+i\alpha_{-1}(x,t)}{2}}W^{(j)}(X),
   \end{array} \end{equation}
where $\delta(t),~\alpha_{\pm1}(x,t)$ are given in (\ref{eqns:ps}), and $U^{(j)},~V^{(j)},~W^{(j)}~(j=1,2)$ are given in (\ref{eqns:sl})-(\ref{eqns:s3}).
The significance of each quantity is: $\gamma(x,t)=\lambda(t) x(\lambda(t) x+2\delta(t))$, coordinate for observing soliton's envelope; $\alpha_1(x,t)$, coordinate for observing soliton's carrier waves;  $\sqrt{\lambda(t)}e^\frac{\delta(t)^2}{2}$, amplitude of the solitons. It is easy to see that $lim_{|x|\rightarrow\infty}\psi_{\pm1,0}(x)=0$ by direct computation, so these two families of
solutions are localized nonlinear wave solutions.

\vskip 12pt
%\section{Dynamics of the localized nonlinear matter wave }
\centerline{\bf III. DYNAMICS OF THE LOCALIZED NONLINEAR MATTER WAVE}

The localized matter waves given by (\ref{eqns:fs}) feature soliton properties: they propagate undistorted and undergo quasielastic collisions.
 According to different choices of $\overline{\omega}$ and $\epsilon$,
 different types of behaviors  can be classified as:

  (a) {\sl Breathing solitons,} when $\epsilon=0$.

  (b) {\sl Quasiperiodic soliton}, when $\epsilon\neq 0$ and the two linear independent solutions $\xi_1,~\xi_2$ belong to the stability domain of (\ref{eqns:me}).

  (c) {\sl Resonant soliton}, when $\epsilon\neq 0$ and  $\xi_1,~\xi_2$  are in the instability domain of (\ref{eqns:me}).

In this section, we will consider the dynamics of the localized nonlinear matter wave and propose how to control them by the external trapping potentials and the space-time inhomogeneous s-wave scattering lengths in future experiments.
  In the following, we take  $^{23}Na$ condensate containing $3\times10^{6}$ atoms and the parameters  are all taken as $c_{-1}=1,~c_0=0.5,~c_1=1,~b_{12}=0.5,~ A=5, ~B=2.$

\vskip 12pt
%\subsection{Breathing solitons}
\centerline{\bf A. Breathing solitons}
Here, we take $\epsilon=0$. Now we study how the space- and time- dependent nonlinearities $g_n$ and $g_s$ control the dynamics of the localized nonlinear waves. In this case, the interactions
$
\label{eqns:g}g_n=-g_s=-\frac{b_{12}}{4c_{-1}^{2}\lambda(t)}e^{-(\lambda(t) x+\delta(t))^{2}}
$
 are all space- and time- dependent, which can be realized by controlling the optically induced Feshbach resonance  or a confinement induced resonance in the real BEC  experiments. The corresponding localized nonlinear wave can be obtained from (\ref{eqns:fs}). These solutions show different features according to the choice of the parameter $\delta(t)$.

{\sl Case 1.}   $\delta(t)=0$. In this case, we take the frequencies  $\omega_x=20 \pi Hz$ and  $\omega_\perp=50 \pi Hz$, the ratio of the confining frequency  $\omega=0.4$. In Fig.1, we show the evolution of density profiles for the one-order wave function $\psi_{1}^{(1)}$ and $\psi_{-1,0}^{(2)}$ with the above parameters. The density profiles for  $\psi_{2}^{(1)}$ and $\psi_{-1,0}^{(1)}$ are the same to that of $\psi_{1}^{(1)}$ and $\psi_{-1,0}^{(2)}$.
It can be observed  that the density wave packets are localized in space and periodically oscillating in time, which are called breathing solitons. Here $\sqrt{\lambda(t)}$ and $1/\lambda(t)$ are the amplitude and width of the matter wave, respectively.  Fig.1 (a) to Fig.1 (c) describe the density profiles of the one-order wave function $\psi_1^{(1)},~\psi_{-1}^{(2)}$ and $\psi_{0}^{(2)}$, respectively. Fig.1 (d)  demonstrates the total density distribution $|\psi_1^{(1)}|^{2}+|\psi_{-1}^{(2)}|^{2}+|\psi_0^{(2)}|^{2}$ for the spinor BECs. Fig.1 (e) demonstrates the amplitude $\sqrt{\lambda(t)}$ and the width $1/\lambda(t)$ of the wave functions. It is observed that the amplitude and the width of the localized nonlinear waves vary periodically  with the increasing time. In all figures of this paper, the units of space length and time are $1.38\mu m$ and $0.7 ms$.

%\vskip 10pt
\begin{figure}[h]
\centering
\includegraphics[width=1.1\textwidth]{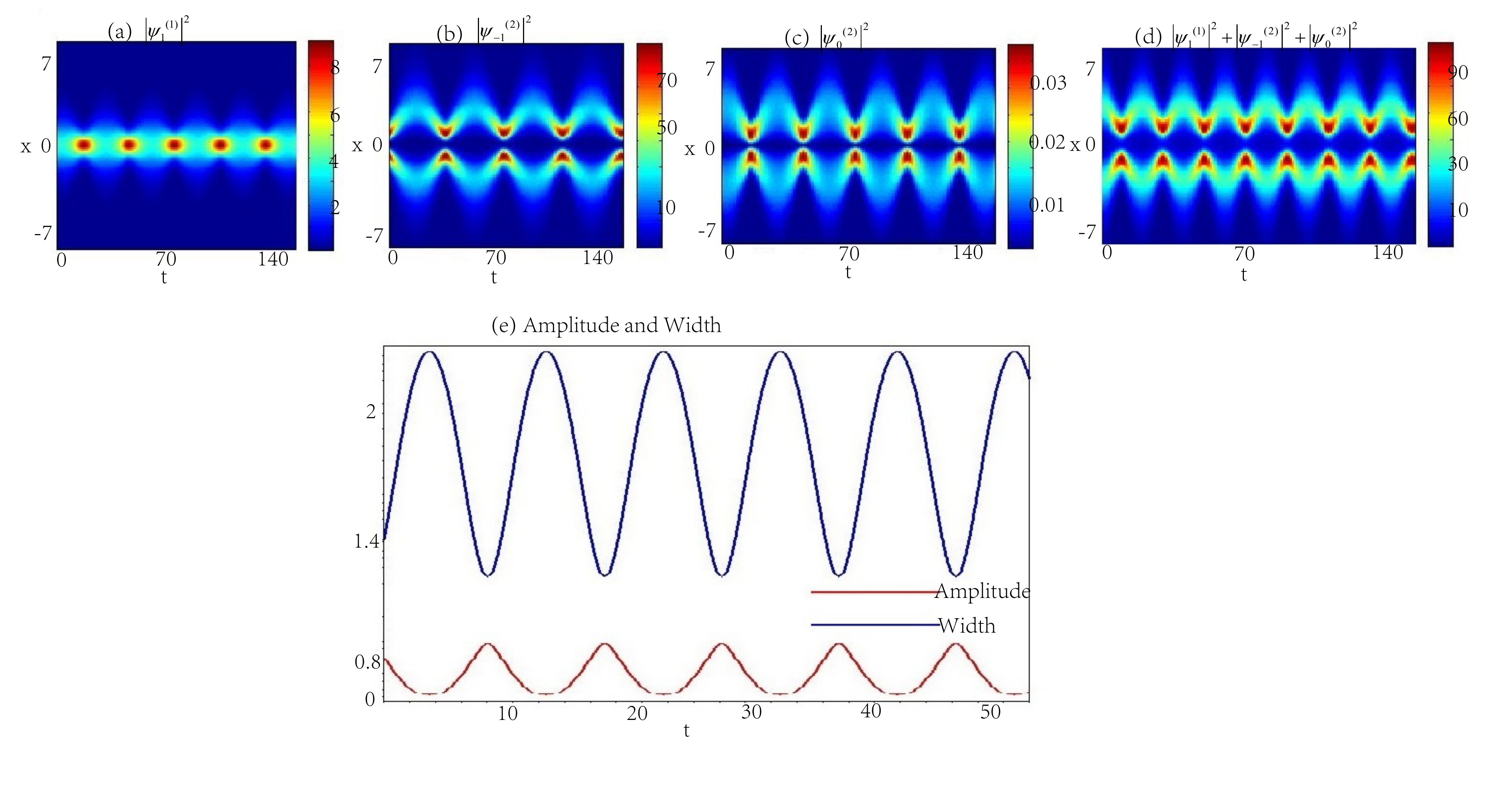}
%\caption{fig1}
%\label{fig:1}
\end{figure}
\begin{center}
\begin{minipage}{ 15cm}{\footnotesize
{\bf FIG. 1.} Dynamics of the breathing solitons in the  spin-1 BEC with spatiotemporally modulated nonlinearities.  (a) to (d)  exhibit the evolution of the density distribution
$|\psi_1^{(1)}|^{2}$, $|\psi_{-1}^{(2)}|^{2}$, $|\psi_0^{(2)}|^{2}$ and the total density distribution
$|\psi_1^{(1)}|^{2}+|\psi_{-1}^{(2)}|^{2}+|\psi_0^{(2)}|^{2}$, respectively.  (e) demonstrates the amplitude (red one) and the width (blue one). The ratio of the confining frequency is taken as $\omega=0.4$.}
\end{minipage}
\end{center}

{\sl Case 2.}   $\delta(t)\neq 0$ and it is given by (\ref{eqns:ps}). In this situation, the tapping potential is still time-independent, but the interactions $g_n,~g_s$ given by  (\ref{eqns:ps}) and  (\ref{eqns:g}) become more complex, the amplitude of the nonlinear matter wave becomes $\sqrt{\lambda(t)} e^{\frac{\delta(t)^2}{2}}$, and the center of mass of the solitons move time periodically  with non-zero velocity. So the nonlinear matter waves are called moving breathing.  For convenience, we assume the ratio of the confining frequencies $\omega$   is time independent to illustrate the dynamics of the moving breathing soliton.  Here we still take $\omega=0.4$ and $\varepsilon=0$.
 Fig. 2 (a)-(c) describe the time evolution of the density profiles of one-order wave function
$\psi_1^{(1)},~\psi_{-1}^{(2)}$ and $\psi_{0}^{(2)}$, respectively.
 Fig. 2 (d)  demonstrates the total density distribution $|\psi_1^{(1)}|^{2}+|\psi_{-1}^{(2)}|^{2}+|\psi_0^{(2)}|^{2}$ for   the spinor BECs. It can be observed the nonlinear matter waves are space localized and moving periodically with respect to time.  Fig. 2 (e) describes  the amplitude of the breathing solitons (red one) and the moving breathing solitons (blue one). It can  be found that the amplitudes of the nonlinear waves vary periodically versus $t$, and the amplitudes of the moving breathing solitons are higher than the breathing solitons.

%\vskip 80pt
\begin{figure}[h]
\centering
\includegraphics[width=1.1\textwidth]{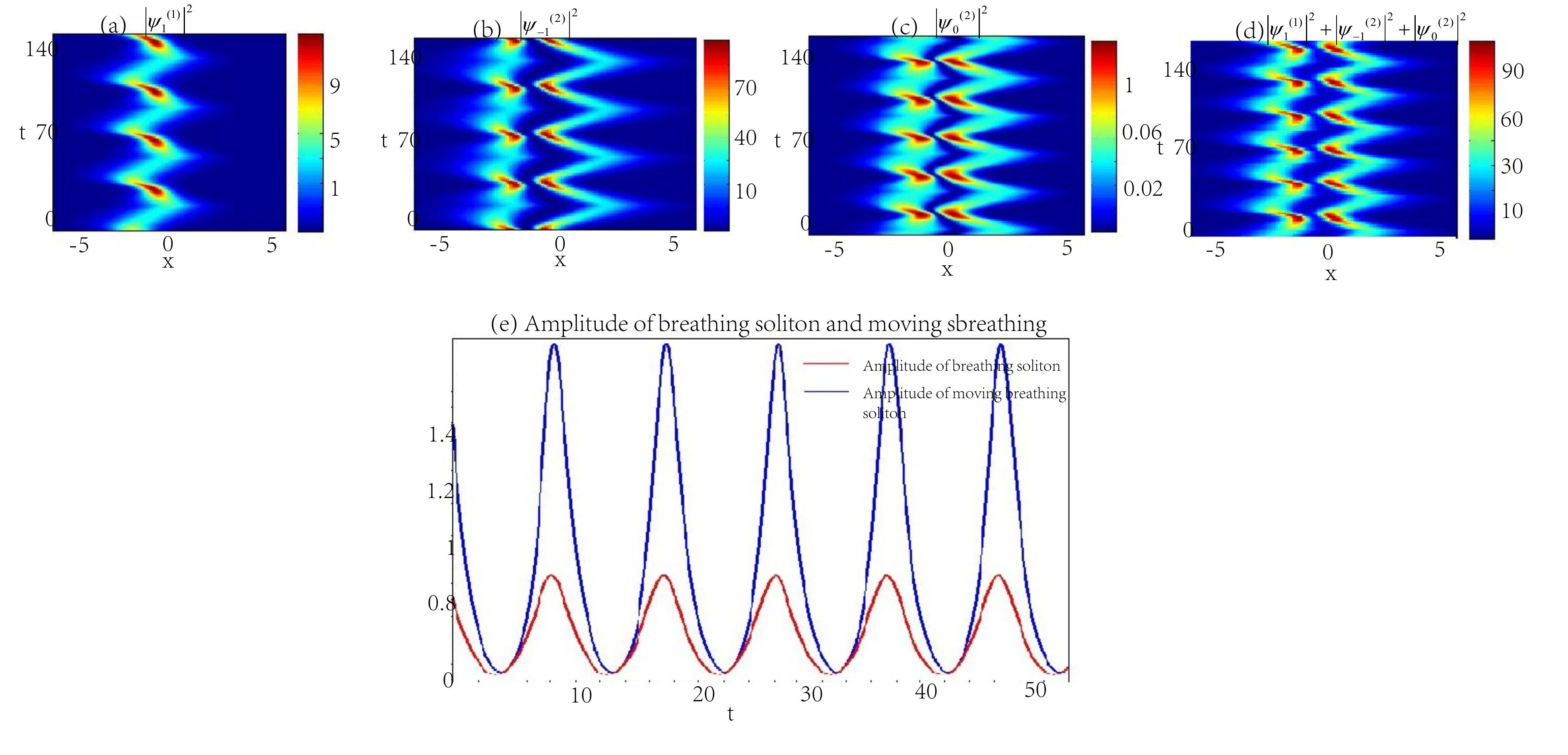}
%\caption{fig1}
%\label{fig:1}
\end{figure}
\begin{center}
\begin{minipage}{ 15cm}{\footnotesize
{\bf FIG. 2.} Dynamics of the moving breathing solitons in the spin-1 BEC with spatiotemporally modulated nonlinearities.  (a) to (d) display the time evolution of the density distribution
$|\psi_1^{(1)}|^2$, $|\psi_{-1}^{(2)}|^2$, $|\psi_0^{(2)}|^{2}$ and  $|\psi_1^{(1)}|^{2}+|\psi_{-1}^{(2)}|^{2}+|\psi_0^{(2)}|^{2}$, respectively. (e) reveals  the amplitude of the breathing solitons (red one) and the moving breathing solitons (blue one). The ratio of the confining frequency is still taken as $\omega=0.4$  and the other parameters are taken as  $c_2=c_3=1/3$ .}
\end{minipage}
\end{center}

 %    For the moving breathing solitons, the parameter $A$ and $B$ play an important role. In Fig.4, we take the wave function $\psi_{-1}$
%as an example to show  how the  parameter $A$ and $B$ effect the dynamics of the localized nonlinear matter wave. When the  parameter $A$ equals to $B$,  %the snakelike soliton appears in Fig.4(a), which maintain the same amplitude in the process of propagation with time. When $A$ and $B$ are different, the %%moving breathing solitons appear and their amplitudes vary periodically versus the time  (Fig. 4(b) and (4c)). Fig. 4(b) demonstrate   the left soliton %travels faster than  the right one when $A<B$. Fig. 4(c) describe the inverse result when $A<B$.

%\vskip 100pt
%\begin{center}
%\begin{picture}(35,30)
%\put(-200,-23){\resizebox{!}{4cm}{\includegraphics{a5b5}}}
%\put(-50,-23){\resizebox{!}{4cm}{\includegraphics{a2b15}}}
%\put(80,-23){\resizebox{!}{4cm}{\includegraphics{a15b2}}}
%\end{picture}
%\end{center}

%\begin{center}
%\begin{minipage}{ 15cm}{\footnotesize ~~~~~~~~~~~~~~~~~~~~~~~~~~~~ (a)~~~~~~~~~~~~~~~~~~~~~~~~~~~~~~(b)~~~~~~~~~~~~~~~~~~~~~~~~~~~~~~~~~~~~~~~(c)\\
%{\bf Figure 4.} Effect of the parameter $A$ and $B$ on  moving breathing solitons.
 %(a) A snakelike soliton  with $\omega_1=0.2, ~\omega_2=0.01, ~c_3=c_4=0.5,~A=B=5$ . (b) The moving breathing soliton with  $\omega_1=0.2, % ~\omega_2=0.01, ~c_3=c_4=0.5,~A=2,~B=14$.  (c) The moving breathing soliton with  $\omega_1=0.2, ~\omega_2=0.01, ~c_3=c_4=0.5,~A=14,~B=2$.}
%\end{minipage}
%\end{center}

\vskip 12pt
%\subsection{Quasibreathing solitons}
\centerline{\bf B. Quasibreathing solitons}
Now, we consider the case of $\epsilon\neq 0$. In order to give an example of quasibreathing solitons, we take $\omega_0=0.4,~\overline{\omega}=2$ and $\epsilon=0.1$ in (\ref{eqns:w}), i.e. the ratio of the confining frequency $\omega=\sqrt{0.16+0.03cos(0.2t) }$ is time dependent function,  which ensure that two linear independent solutions $\xi_1$ and $\xi_2$ of the Mathieu equation (\ref{eqns:w}) belong to its stability region. So  (\ref{eqns:w}) has two incommensurable frequencies. In this way, the localized matter wave given by the solutions (\ref{eqns:fs}) exhibit quasiperiodic behaviors and the trapping potential and interactions are still space and time dependent.
%\vskip 350pt
\begin{figure}[h]
\centering
\includegraphics[width=1.1\textwidth]{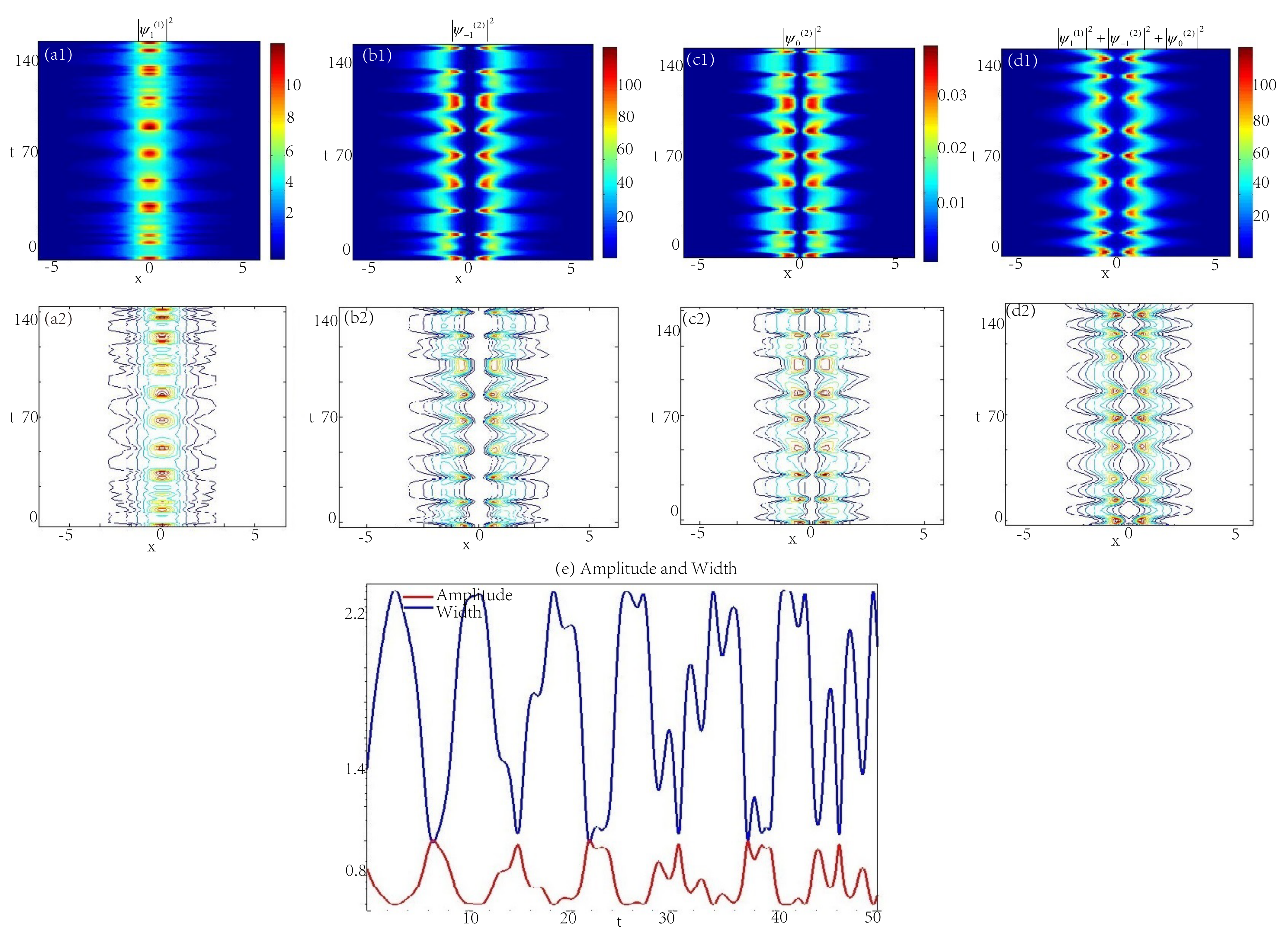}
%\caption{fig1}
%\label{fig:1}
\end{figure}
\begin{center}
\begin{minipage}{ 15cm}{\footnotesize
{\bf FIG. 3.} Dynamics of the quasibreathing  solitons in the  spin-1 BEC with spatiotemporally modulated nonlinearities.  (a1) and (a2) express  the evolution of the density and contour distribution
$|\psi_1^{(1)}|^{2}$, respectively. (b1) and (b2) exhibit the evolution of the density and contour distribution
$|\psi_{-1}^{(2)}|^{2}$, respectively.(c1) and (c2) display the evolution of the density and contour distribution
$|\psi_0^{(2)}|^{2}$, respectively. (d1)  and (d2) illustrate the evolution of the density and contour distribution
$|\psi_1^{(1)}|^{2}+|\psi_{-1}^{(2)}|^{2}+|\psi_0^{(2)}|^{2}$. (e) demonstrates the amplitude (red one) and the width (blue one). The parameters are taken as $\omega_1=0.03cos(2t), ~\omega_2=0.2, ~\delta=0$. }
\end{minipage}
\end{center}
Fig. 3 shows an example of the  quasiperiodic behavior. The first column to the fourth column describe the time evolution of the density and contour profiles of the one-order wave function $\psi_1^{(1)}$, $\psi_{-1}^{(2)}$, $\psi_0^{(2)}$ and the total density distribution $|\psi_1^{(1)}|^{2}+|\psi_{-1}^{(2)}|^{2}+|\psi_0^{(2)}|^{2}$, respectively.
%Fig. 3(a1) and (a2) demonstrate the evolution of the density and contour profiles of the one-order wave function $\psi_1^{(1)}$, respectively. Fig. 3(b1) and (b2) describe the evolution of the density and contour profiles of wave functions
%$\psi_{-1}^{(2)}$, respectively. Fig. 3(c1) and (c2) describe the evolution of the density and contour  profiles of the one-order wave function
%$\psi_0^{(2)}$, respectively. Fig. 3(d1) and (d2) demonstrates the density and contour distribution $|\psi_1^{(1)}|^{2}+|\psi_{-1}^{(2)}|^{2}+|\psi_0^{(2)}|^{2}$.
Fig. 3 (e) describes  the amplitude  (red one) and the width (blue one) of the quasibreathing solitons. It can be observed the nonlinear matter waves are space localized and quasiperiodic with respect to time and it can  also be fond that the amplitude and width of the nonlinear waves are quasiperiodically versus $t$.

\vskip 12pt
%\subsection{Resonant solitons}
\centerline{\bf C. Resonant solitons }

Here, we still consider the case of $\epsilon\neq 0$, i. e.,  the ratio of the  confining frequency $\omega$ is time dependent. In this case, we choose $\omega_0=0.44,~\overline{\omega}=32$ and $\epsilon=0.003$ in Mathieu equation (\ref{eqns:w}), which ensure that two linear independent solutions $\xi_1$ and $\xi_2$ of (\ref{eqns:w}) belong to its instability region. Thus, the localized matter waves given by the  solutions (\ref{eqns:fs}) show the resonant solitons behaviors and the trapping potential and interactions are still space and time dependent.
In Fig. 4, we show the evolution of the density  and contour profiles for the  resonant solitons. The first to fourth column in Fig.4 demonstrate the evolution of the density and contour  profiles of the one order wave function
$\psi_1^{(1)},~\psi_{-1}^{(2)}$, $\psi_{0}^{(2)}$,  and $|\psi_1^{(1)}|^{2}+|\psi_{-1}^{(2)}|^{2}+|\psi_0^{(2)}|^{2}$, respectively. It can be observed the nonlinear matter waves are space localized and time resonant. Fig. 4 (e) shows the amplitude and width versus time for the resonant  solitons.    It can be observed  that the amplitude of the resonant solitons  is low   and its width are large at beginning. After a while, the amplitudes become higher but the widths become small. This phenomenon appear gradually as time goes. So this nonlinear matter wave demonstrate the resonant soliton behavior. This resonant behaviors comes from the cooperation of the  spatiotemporal inhomogeneous interactions and trapping potential.

\vskip 75pt
\begin{figure}[h]
\centering
\includegraphics[width=1.1\textwidth]{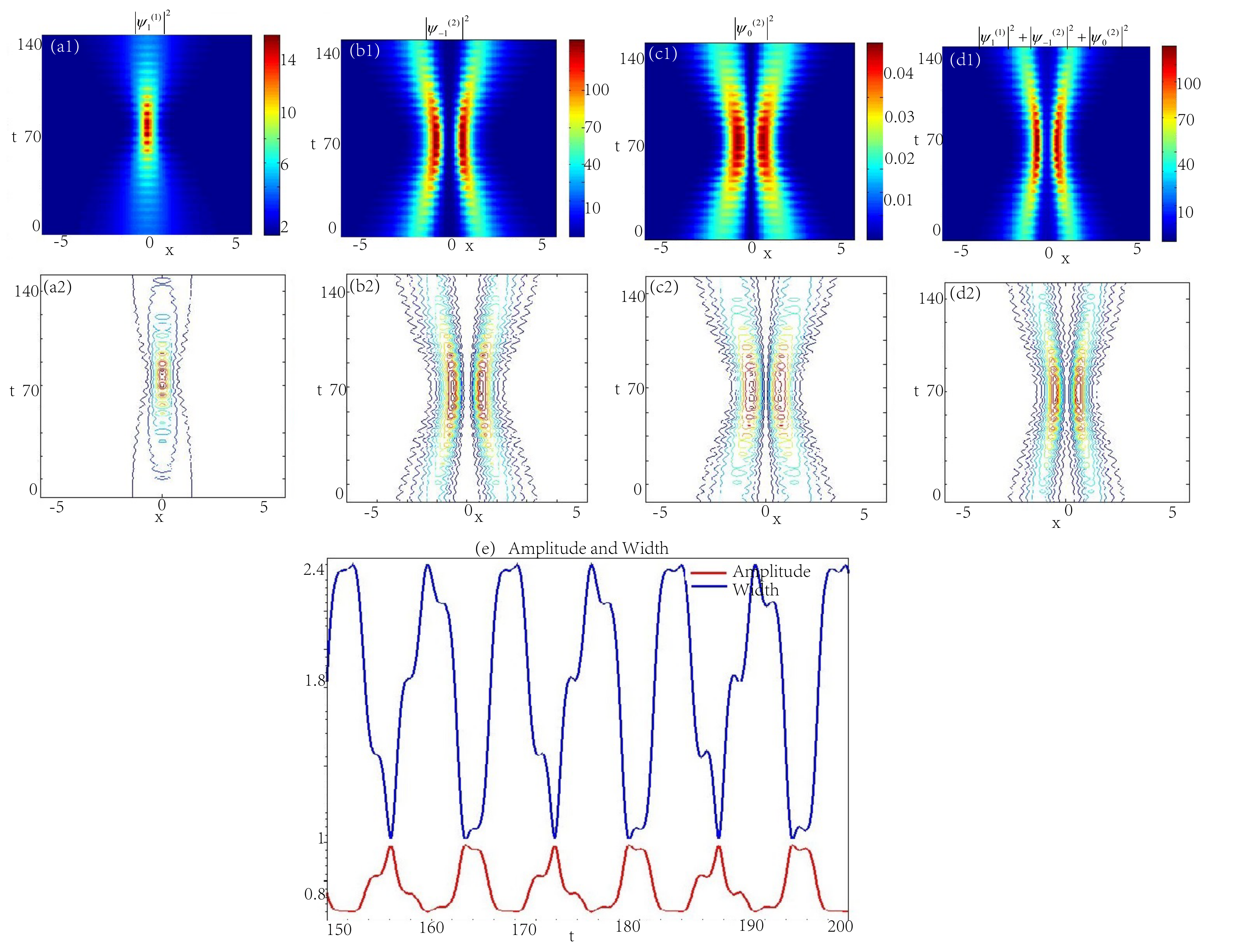}
%\caption{fig1}
%\label{fig:1}
\end{figure}
\begin{center}
\begin{minipage}{ 15cm}{\footnotesize
{\bf FIG. 4.} Dynamics of the Resonant breathing solitons in the  spin-1 BEC with spatiotemporally modulated nonlinearities.  (a1) and (a2) explain the evolution of the density and contour
$|\psi_1|^{2}$, respectively. (b1) and (b2) exhibit the evolution of the density and contour
$|\psi_-1|^{2}$, respectively. (c1) and (c2) express the evolution of the density and contour
$|\psi_0|^{2}$, respectively. (d1) and (d2) demonstrate the evolution of the density and contour
$|\psi_1^{(1)}|^{2}+|\psi_{-1}^{(2)}|^{2}+|\psi_0^{(2)}|^{2}$, respectively. (e) shows  the amplitude of the breathing solitons (red one) and the moving breathing solitons (blue one). Here the ratio of the confining frequency $\omega=\sqrt{0.19+0.03 cos32t}$.  }
\end{minipage}
\end{center}

\vskip 12pt
%\section{Stability analysis}
\centerline{\bf IV. STABILITY ANALYSIS}
Now we study the dynamical stability of the localized nonlinear wave solutions  (\ref{eqns:fs}) by performing some numerical simulations. Here we run the numerical simulations by use of the split-step Fourier transformation. The domain is compose of $N=512$ grids point and the step sized of time integration is $\tau=0.0001$. We take $\psi_{j}(x,0)~(j=-1,~0,~1)$ as an initial values and the simulations are lasting up to $t=200$. The simulation results show that:

 (a) When $\delta=0$, the exact localized nonlinear wave solutions  (\ref{eqns:fs}) is dynamically stable for $j=1$ and $n=0$, that is to say the one order breathing soliton, quasibreathing soliton and resonant soliton are all stable.

 (b) When $\delta\neq 0$, the moving solitons $\psi^{(1)}_{1}$ and $\psi^{(1)}_{0}$ are dynamically stable, while $\psi^{(1)}_{-1}$ is unstable.

Fig. 5 show the time evolution of the one-order breathing solitons for  $\delta=0$ and $\varepsilon=0$ with $g_n(x,t)=-g_s(x,t)=\frac{1}{8}\nu(t)^{-\frac{1}{2}}e^{-\frac{x^{2}}{\nu(t)}}
 ,~\nu(t)=5sin(\frac{2}{5}t)^2+2cos(\frac{2}{5}t)^2+\sqrt{15}sin(\frac{2}{5}t)cos(\frac{2}{5}t)$. It can be observed that the one-order breathing solitons $\psi^{(1)}_{1}$ ,  $\psi^{(1)}_{0}$  and $\psi^{(1)}_{-1}$  are dynamical stable.

\begin{figure}[h]
\centering
\includegraphics[width=1\textwidth]{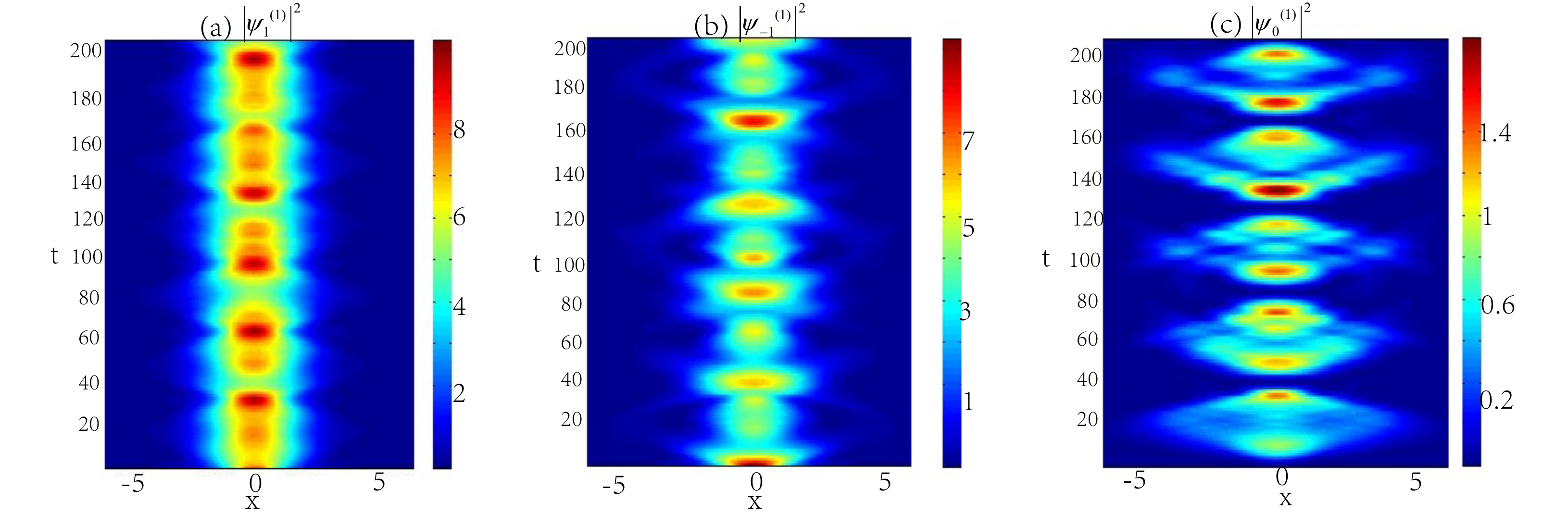}
%\caption{fig1}
%\label{fig:1}
\end{figure}
\begin{center}
\begin{minipage}{ 15cm}{\footnotesize
{\bf FIG. 5.} Evolution of the one-order  breathing solitons  for  $\delta=0$, $\varepsilon=0$  and the nonlinearity $g_n(x,t)=-g_s(x,t)=\frac{1}{8}\nu(t)^{-\frac{1}{2}}e^{-\frac{x^{2}}{\nu(t)}}$.  The other parameters are the same as used in Fig1.}
\end{minipage}
\end{center}

Fig.6 show the evolution of the one-order  moving breathing solitons  with  $\delta=1$ , $\varepsilon=0$  and $g_n(x,t)=-g_s(x,t)=\frac{1}{8}\nu(t)^{-\frac{1}{2}}e^{(1-\frac{x}{\nu(t)})^2}$. It can be seen  that the moving  breathing solitons
  $\psi_1^{(1)}$ and $\psi_0^{(1)}$  are dynamical stable, while $\psi_{-1}^{(1)}$ is unstable.
\vskip 70pt
\begin{figure}[h]
\centering
\includegraphics[width=1\textwidth]{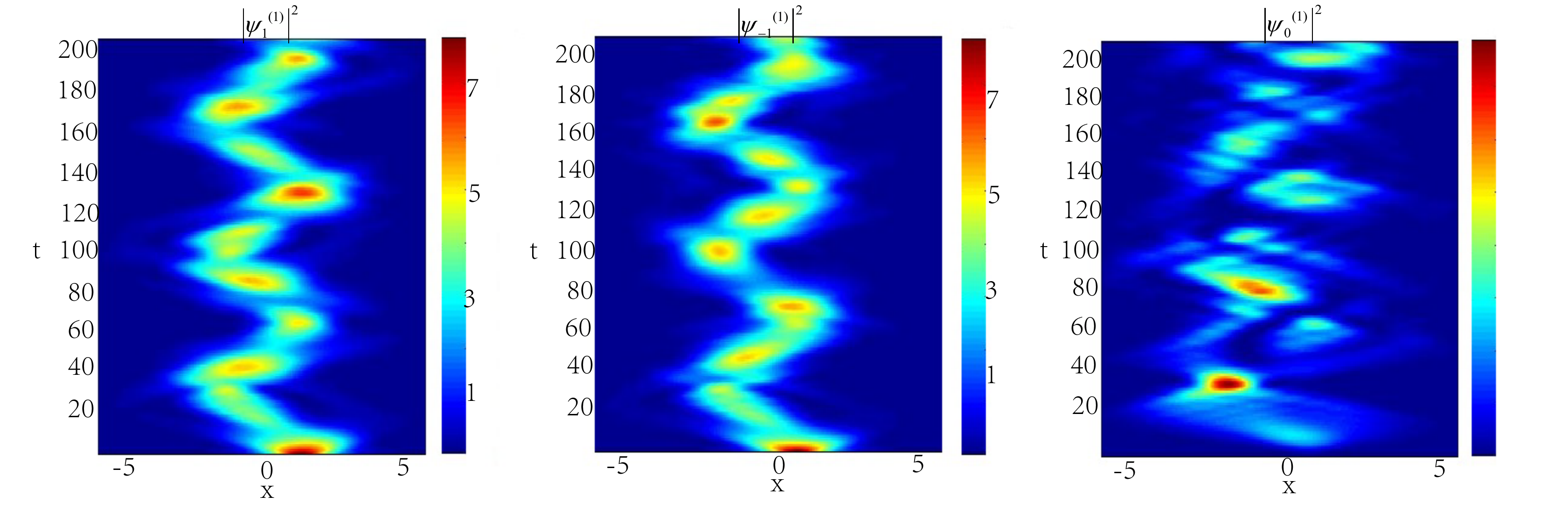}
%\caption{fig1}
%\label{fig:1}
\end{figure}
\begin{center}
\begin{minipage}{ 15cm}{\footnotesize
{\bf FIG. 6.} Evolution of the one-order  moving breathing solitons   with $\delta=1$, $\varepsilon=0$  and the nonlinearity $g_n(x,t)=-g_s(x,t)=\frac{1}{8}\nu(t)^{-\frac{1}{2}}e^{(1-\frac{x}{\nu(t)})^2}$.  The other parameters are the same as used in Fig2.}
\end{minipage}
\end{center}

\vskip 12pt
%\section*{Discussion}
\centerline{\bf V. CONCLUSION}
In this paper, we  study the dynamics of the localized nonlinear matter wave solutions of the three-component GPEs  with time- and space- dependent nonlinearities
for F=1 spinor BECs. These solutions are derived  in terms of Jacobi elliptic functions and Mathieu equation by similarity transformation.
Further, we  illustrate that the localization of the nonlinear matter wave takes the form of  the vector
breathing solitons, moving breathing solitons, quasibreathing solitons and resonant solitons. The dynamical stability of the all kinds of vector solitons
are analysed by numerical stimulation.   The results show that the one-order breathing solitons, quasibreathing solitons, resonant soltons and the moving breathing solitons except for the  matter wave $\psi_0$ are all stable.

We take the sodium atom $^{23}Na$  with the total number  $N=3\times10^{6}$ as an example to show how to create various soliton phenomena under the condition that the Zeeman energy $E_j$ satisfies $E_1+E_{-1}=E_0$.  When the confining frequencies in the transverse and axial directions are taken as   $\omega_x=20 \pi Hz$ and  $\omega_\perp=50 \pi Hz$, respectively, the breathing solitons can be observed with the interactions   $g_n(x,t)=-g_s(x,t)=\frac{1}{8}\nu(t)^{-\frac{1}{2}}e^{-\frac{x^{2}}{\nu(t)}}$, the moving breathing solitons can be observed when $g_n(x,t)=-g_s(x,t)=\frac{1}{8}\nu(t)^{-\frac{1}{2}}e^{(1-\frac{x}{\nu(t)})^2}$. When the confining frequencies in the transverse are the functions of t, the quasibreathing soliton and resonant solitons may be observed. For example, the interactions $g_n(x,t)=-g_s(x,t)=-e^{(\lambda(t) x+\delta(t))^2}/(8\lambda(t))$ and the confining frequencies $\omega_x=(16+3 cos0.2t) \pi Hz$ and  $\omega_\perp=100 \pi Hz$,  the quasibreathing soliton can appear. And the resonant soliton can appear with  the confining frequencies $\omega_x=(19+0.03 cos32t) \pi Hz$ and  $\omega_\perp=100 \pi Hz$.
  We hope that these dynamics behaviors of the spin-1 BECs  with spatiotemporal  nonlinearities  can be realized in the future experiment and help us to understand these phenomena further.

\section*{Acknowledgements}
This work was supported by the NKRDP under grants Nos. 2016YFA0301500, NSFC under grants Nos. 11434015, 61227902, 61378017, KZ201610005011, SKLQOQOD under grants No. KF201403, SPRPCAS under grants No. XDB01020300, XDB21030300.

\end{document}